\documentclass[runningheads]{llncs}
\usepackage{graphicx}
\usepackage{pgfplots}
\usepackage{wrapfig}
\usepgfplotslibrary{statistics}
\usepackage{bibnames}
\usepackage[]{algorithm2e}
\RestyleAlgo{boxruled}
\LinesNumbered
\graphicspath{{images/}}
\newcommand{\comment}[1]{}

\begin{document}
%
\title{Fault-Aware Non-Collective Communication Creation and Reparation in MPI}
%
\titlerunning{Combining Fault-Awareness and Non-Collectiveness in MPI}
%
\author{Roberto Rocco, Gianluca Palermo}
%
\authorrunning{R. Rocco and G. Palermo}
%
\institute{Politecnico di Milano\\Dipartimento di Elettronica e Informazione, Italy \\
\email{\{name.surname\}@polimi.it}}
\maketitle              
\begin{abstract}
The increasing size of HPC architectures makes the faults' presence a more and more frequent eventuality. This issue becomes especially relevant since MPI, the de-facto standard for inter-process communication, lacks proper fault management functionalities. Past efforts produced extensions to the MPI standard that enabled fault management, including ULFM. While providing powerful tools to handle faults, it still faces limitations like the collectiveness of the repair procedure. With this paper, we overcome those limitations and achieve fault-aware non-collective communicator creation and reparation. We integrate our solution into an existing fault resiliency framework and measure the overhead introduced in the application code. The experimental campaign shows that our solution is scalable and introduces a limited overhead, and the non-collective reparation is a viable opportunity for ULFM-based applications.

\keywords{Fault Management \and MPI \and ULFM}
\end{abstract}
\section{Introduction}\label{sec:intro}

Computational science applications require more and more resources for their computation, leading to the growth of current HPC systems in terms of performance, energy efficiency and complexity. HPC systems have recently reached the exascale boundary (10\textsuperscript{18} FLOPS) \cite{Top500}, but the additional performance brings new issues. The increase in the number of nodes brought a greater probability of faults, and HPC systems (and their applications) must be able to handle them. Studies \cite{egwutuoha2013survey} show that the impact of faults is already relevant in HPC. This result also comes from the absence of fault management techniques in the Message Passing Interface (MPI) \cite{clarke1994mpi}, the de-facto standard for inter-process communication. The last version of the MPI standard (4.0) tried to reduce the possible sources of faults, isolating them into single processes when possible. While removing the faults' impact on local functions, it does not avoid their propagation with communication and does not provide a way to repair the execution.

The issue of fault tolerance in MPI is not new to the field and led to the production of fault-aware MPI implementations \cite{bouteiller2006mpich,ferreira2011rmpi,fagg2000ft}. Most of them received limited support and did not solve the problem entirely and efficiently. The User-Level Fault Mitigation (ULFM) MPI extension \cite{bland2013post} is currently one of the most relevant works in this direction. ULFM features a collection of functions for fault detection and notification, structure reparation, and execution resiliency. Among the added functionalities of ULFM, the \textit{shrink}, \textit{agree}, and \textit{revoke} functions enable communicator reparation, resilient agreement, and fault propagation, respectively. ULFM is currently integrated directly into the latest versions of OpenMPI, one of the most relevant MPI implementations. 

While ULFM provided powerful new possibilities for fault management to users, its introduction inside the application code is non-trivial. For this reason, many efforts adopted the functionalities introduced by ULFM to produce frameworks able to provide checkpoint and restart (C/R) with automatic rollback \cite{gamell2014exploring,losada2017resilient,teranishi2014toward}. Others leveraged those capabilities to isolate the failed process from the execution, limiting the effect of the fault \cite{pauli2015intrinsic,rocco2021legio}. The idea behind these frameworks is to simplify the introduction of fault tolerance functionalities inside applications. Nonetheless, ULFM and all the frameworks based on it still face a strong constraint: the reparation process is always collective. While this constraint is negligible in most cases, it could be a limitation when dealing with certain MPI functionalities, like non-collective calls.

In this effort, we overcome the ULFM repair collectiveness constraint to introduce fault handling in applications using non-collective calls.
In particular, we focus on applications using non-collective communicator creation functions (\texttt{MPI\_Comm\_create\_group} and \texttt{MPI\_Comm\_create\_from\_group}).
Previous works have already shown the relevance of the first call \cite{dinan2011noncollective}, which has been in the MPI standard since version 3.0. The other function is a recent addition (version 4.0) fundamental to the session execution model.
We propose a Liveness Discovery Algorithm to overcome the collective reparation constraint, avoiding unnecessary synchronizations and allowing non-collective communicator creation, even with faults among the caller processes.

The contributions of this paper are the following:
\begin{itemize}
\item We analyze the effects of faults on the two non-collective communicator creation calls \texttt{MPI\_Comm\_create\_group} and \texttt{MPI\_Comm\_create\_from\_group};
\item We design and implement a Liveness Discovery Algorithm to non-collectively detect the failed processes and limit their impact on the non-collective communicator creation functions;
\item We use the Liveness Discovery Algorithm to reimplement two of the most important ULFM functionalities with non-collective behaviour.
\item We integrate the solution into an existing fault resiliency framework to simplify its usage inside the user code, and we evaluate the introduced overhead.
\end{itemize}

The paper is structured as follows: Section~\ref{sec:background} discusses ULFM and the frameworks that leverage its functionalities.
Section~\ref{sec:non_collective} illustrates the behaviour of the two non-collective operations analyzed. Section~\ref{sec:tree} discusses the Liveness Discovery Algorithm, its integration and the possibilities it enables. Section~\ref{sec:experimental} covers the experimental campaign done to evaluate the proposed solution overhead and scalability. Lastly, Section~\ref{sec:conclusion} concludes the paper.

\section{Background and previous work}\label{sec:background}

The failure of an MPI process can impact an application execution since the MPI standard does not specify the behaviour of the survivor processes. The last version of the MPI standard (4.0) introduced new functionalities to simplify fault handling, limiting their impact when possible. While providing ways to represent and react to faults, the standard still does not contain a defined method to recover the execution. After a fault occurrence, the best outcome planned by the MPI standard is the graceful termination of the application.

One of the most relevant efforts that provide tools for the continuation of the execution is ULFM \cite{bland2013post}. It is an MPI standard extension proposal focusing on fault detection, propagation and reparation. While still under development, the ULFM extension got included in the latest versions of OpenMPI, one of the principal MPI implementations. The idea behind ULFM is to allow application developers to manage faults by themselves. The user can change the application code by introducing ULFM functions directly. This approach allows maximum flexibility in fault management at the cost of additional integration complexity: the programmer must know how and when to handle faults, which is non-trivial.

The latest developments of the ULFM extension \cite{bouteiller2022implicit} introduced new functionalities that can simplify the interaction between the application and the fault tolerance functionalities. With the use of \texttt{MPI\_Info} structures, the user can specify the error propagation policy and automate the failure notification phase after fault detection. They also included the non-blocking communicator repair functionality: it removes the need for coordination in case of multiple reparations and enables overlaps between application-level recovery and communicator reparation.

While introducing new functionalities to allow the execution past the rise of a fault, ULFM does not provide any mechanism to recover the execution. This decision comes from the fact that different applications may require different types of recovery, while some do not. The user should choose the best recovery mechanism and integrate it directly into its code. This approach gives maximum flexibility to the user but introduces unneeded complexity in the user code. For this reason, many efforts produced all-in-one frameworks that include ULFM and a recovery mechanism to simplify fault management integration inside user code \cite{gamell2014exploring,shahzad2018craft,teranishi2014toward,gamell2015local,losada2017resilient,losada2019local,suo2013nr,rocco2021legio}. 

The principal research efforts in this direction are towards ULFM and Checkpoint/Restart (C/R) functionalities integration \cite{gamell2014exploring,shahzad2018craft,teranishi2014toward,gamell2015local,losada2017resilient,losada2019local,suo2013nr}. This approach introduces fault tolerance (the ability to nullify the effect of a fault) in generic MPI applications. All these efforts came to similar solutions, with the execution restarting from the last consistent state, removing the fault impact. This approach simplifies the introduction of fault tolerance in MPI applications since it hides the complexity within the framework. Moreover, some efforts \cite{losada2017resilient,losada2019local} removed the need for code changes in the application by leveraging a heuristic code analysis to choose the best integration with their framework.

The solution adopted in Legio \cite{rocco2021legio} is slightly different: the effort produced a library that introduces fault resiliency (the ability to overcome a fault) in embarrassingly parallel applications. Applications using Legio continue after the fault detection, but the failed processes will not resume: the execution proceeds only with the survivor processes, causing a loss of correctness but resuming the execution faster. The authors claim that these characteristics make Legio ideal for approximate computing applications, where the algorithms already trade correctness for speed. Legio follows the policy of transparent integration with the application: it does not require any code change.

To summarize, many efforts base on ULFM to manage faults in MPI applications. ULFM is evolving, introducing support for non-blocking repairs and different fault notification policies \cite{bouteiller2022implicit}. The focus of the new additions is to allow the reparation to happen while performing the computations needed for the execution continuation. While these improvements are welcome, they still require eventual participation from all the processes and are thus orthogonal to non-collectiveness. This assumption blocks the users from gaining the benefits of non-collective communication creation, resulting in a potential loss of efficiency. In this work, we find a Liveness Discovery Algorithm that handles faults non-collectively and removes the collectiveness constraint from ULFM functionalities.

\section{Non-collective operations}\label{sec:non_collective}

A non-collective operation involves many processes inside an MPI communicator, but not necessarily all. In the MPI standard, it is possible to find a few examples of non-collective calls, like the \texttt{MPI\_Comm\_create\_group} call. First proposed in \cite{dinan2011noncollective}, it got introduced with version 3.0 of the standard. The function creates a communicator containing only the processes part of the group structure passed as a parameter. This function must be called only by the group participants, not by all the processes in the communicator (differently from the \texttt{MPI\_Comm\_create} function). The call \texttt{MPI\_Comm\_create\_from\_group} behaves similarly but does not require a starting communicator. Introduced with version 4.0 of the standard, it is part of the functions that handle the session execution model. It allows the creation of a communicator starting only from a group of processes. The absence of a parent communicator makes this call unique and potentially problematic in case of faults.

The non-collectiveness of the two calls poses some additional difficulties. The ULFM solution consists of letting the function raise an error, agreeing on the correctness of the operation (with the \textit{agree} function) and, if any processes show an error, repairing the communicator (with the \textit{shrink} function) and retry. The two ULFM calls mentioned are, however, collective. Using these functions may require support from processes not involved in the communicator creation. Forcing those processes to collaborate loses the benefits of non-collective communicator creation (less synchronization), so it is not feasible. To introduce fault management in the non-collective call, we would ideally not use any collective operation in general. With this restriction, it is only possible to notify the fault presence (using the \textit{revoke} call) to the other processes not involved in the communication creation. While this allows for an eventual reparation, its collectiveness precludes any benefit from non-collective call usage.

We conducted some preliminary experiments to evaluate the behaviour of non-collective calls with faults. To better represent the fault occurrence in a communicator, we describe it as either \textit{faulty} or \textit{failed}.
In particular, \textit{faulty} communicators contain some failed processes with no process acknowledging them. When a process discovers the failure, the communicator becomes \textit{failed}, and the failure propagation begins. We ran those experiments using the latest version of OpenMPI featuring ULFM (v5.0.0) and implementing the standard MPI 4.0. Our tests on the function \texttt{MPI\_Comm\_create\_group} proved that:
\begin{itemize}
	\item The call works if the communicator passed as a parameter is \textit{faulty} as long as no process part of the group failed;
	\item The call deadlocks with a \textit{faulty} communicator parameter if one process part of the group failed;
	\item The function fails with a \textit{failed} communicator as a parameter, returning the ULFM-defined error code \texttt{MPIX\_ERR\_PROC\_FAILED}, regardless of the presence of failed processes inside the group.
\end{itemize}

While we think that the deadlock eventuality is an implementation flaw of the current ULFM version, it is still a challenge to remove. ULFM should control the presence of failed processes within the group, even the ones with unacknowledged failures. However, ULFM does not specify any function to perform proactive fault management (controls before raising an error). So ULFM does not provide any call to handle this case in a non-collective way.

Our tests on the function \texttt{MPI\_Comm\_create\_from\_group} produced similar results: the call works if failed processes are not part of the group passed as a parameter, and it deadlocks otherwise. A fix for the first function can also solve the deadlock eventuality on this one. Therefore, this effort aims to create an algorithm for the remotion of failed processes from the group parameters of the non-collective functions.

\section{Liveness Discovery Algorithm}\label{sec:tree}

The problem of finding which processes failed within a group is dual to the one of discovering which are still alive. In a collective scenario, we can solve the latter using the \texttt{MPI\_Allgather} function: each process can share its rank with the others and obtain data about all the processes sharing. If an error arises, all the participants can share a communal view using the \textit{agree} functionality and then proceed to remove the failures from the communication with the \textit{shrink} call. The execution can repeat these steps until the function \texttt{MPI\_Allgather} completes correctly: after the completion, each process has the list of survivor ranks since all and only the survivors can communicate with no errors in a collective call.

This solution faces many problems when moving to the non-collective scenario. All the functions above are collective and not usable for fault-checking non-collective calls. While we cannot use operations like the \texttt{MPI\_Allgather}, we can follow the same data movement: the idea is to let processes exchange data in an all-gather pattern but implement it with point-to-point communication. Moreover, the solution must be resilient to process faults, so a standard algorithm may not be sufficient.

\begin{figure}[!t]
    \centering
    \includegraphics[width=0.5\textwidth]{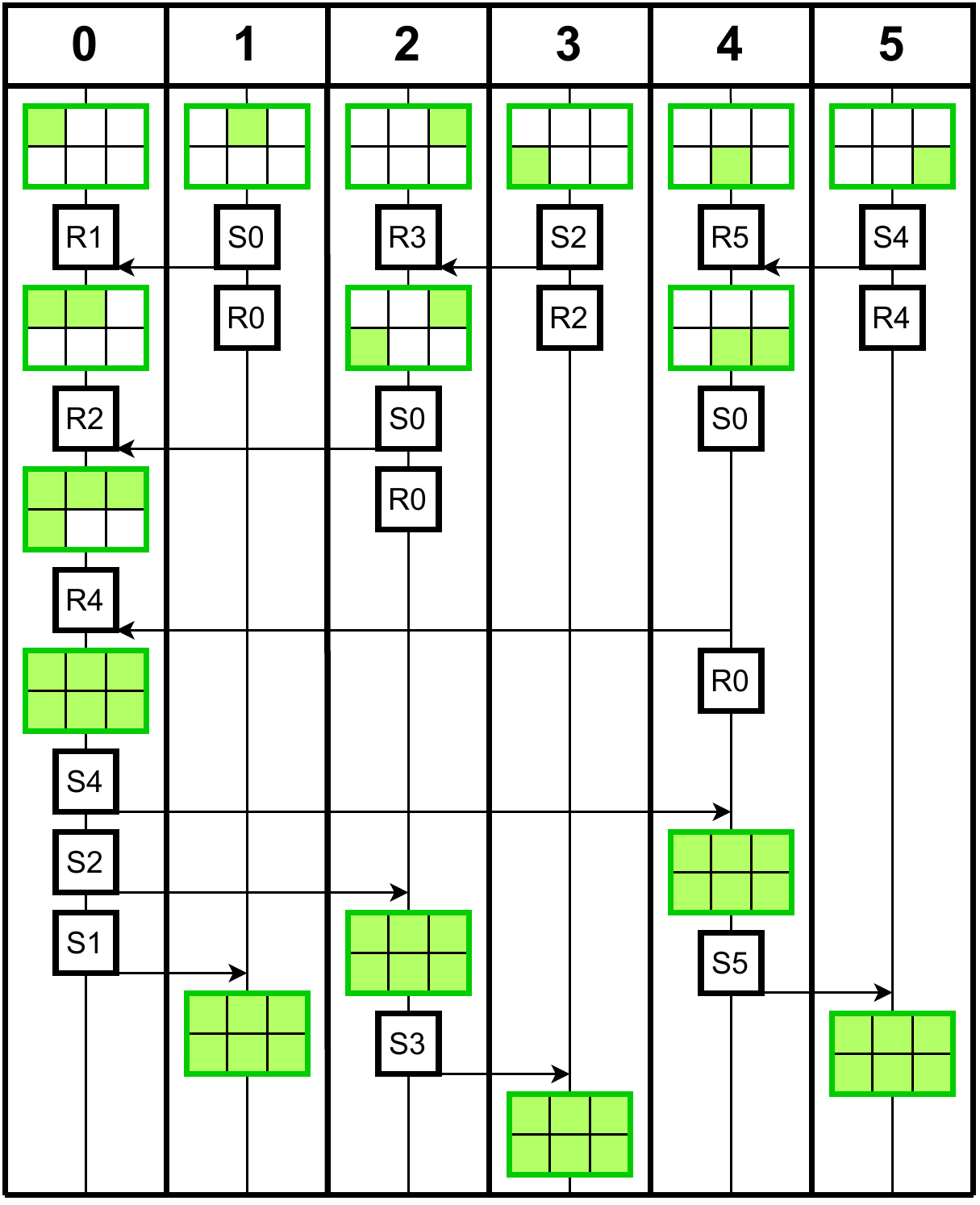}
    \caption{A representation of the Liveness Discovery Algorithm. Each column represents a process, while green rectangles contain the data on the liveness of each participant. Black boxes represent MPI operations, with the first character showing the type (\textbf{S}end/\textbf{R}ecv) and the second the rank of the other process.}
    \label{fig:tree}
\end{figure}

\begin{algorithm}[!b]
    \caption{Naive version of the Liveness Discovery Algorithm}
    \label{alg:tree}
    \SetAlgoVlined
    \KwIn{A processes' group of size \texttt{s}, each process with rank \texttt{r} (from 0 to \texttt{s}-1)}
    \KwOut{A processes' group containing only non-failed processes}
    \tcc{all ranks are encoded using the minimum number of bits}
    data.append(r)\tcp*[l]{data contains only own rank}
    root\_level = \textit{number of trailing zeros of \texttt{r}}\;
    root\_index = 1\;
    \While{root\_index $\leq$ root\_level}{
        partner = r + (1 $<<$ root\_index)\tcp*[l]{receive from rank far from root}
        \If{partner $<$ s} {
            \textit{receive data from partner and append to known}
        }
        root\_index++\;
    }
    partner = r - (1 $<<$ root\_index)\tcp*[l]{send data to rank closet to root}
    \If{root\_index $<$ \textit{bits used for encoding}}{
        \textit{send all data to partner\\}
        \textit{receive full data from partner, substitute own data with the received one}
    }
    \tcc{Start the propagation towards leaves}
    root\_index--\;
    \While{root\_index $>$ 0} {
        partner = r + (1 $<<$ root\_index)\;
        \If{partner $<$ s} {
            \textit{send full data to partner}
        }
    }
    \tcc{Now all the processes' data contains all the non-failed ranks}
\end{algorithm}

The algorithm we propose is the \texttt{MPI\_Allgather} recreation as a combination of gather and broadcast operations. When collective calls are not feasible, applications mostly use point-to-point operations in a tree-shaped communication pattern to implement the gather and broadcast functionalities. Following this concept, we designed and implemented an algorithm that executes the gather and broadcast phases. Algorithm~\ref{alg:tree} contains a pseudo-code implementation of this first design, while Figure~\ref{fig:tree} shows a sample execution with six processes. 

\begin{figure}[!b]
    \centering
    \includegraphics[width=0.5\textwidth]{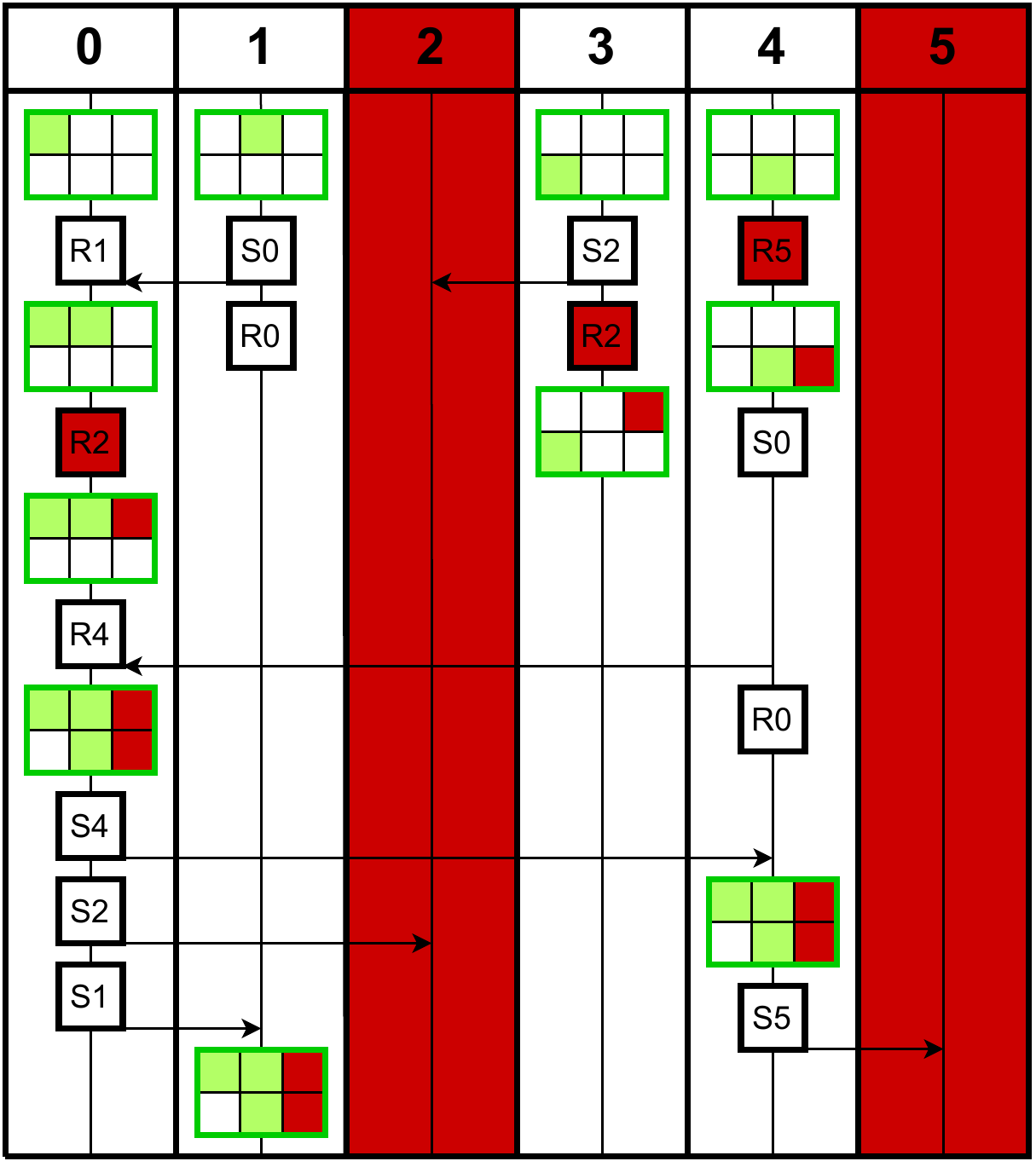}
    \caption{The figure shows the algorithm depicted in Figure~\ref{fig:tree} in presence of faults. Red rectangles represent failed processes and MPI calls.}
    \label{fig:tree_problem}
\end{figure}

While this solution works in a fault-free scenario, failed processes' presence can compromise the result's correctness. This eventuality is not due to error propagation, MPI fault management or additional ULFM functionalities but rather an algorithmic issue. Each rank value has a single path towards all the other nodes: if it breaks, the information will not arrive since no fallback strategy is present. Figure~\ref{fig:tree_problem} shows the erroneous behaviour in execution with two faults: the algorithm does not correctly behave since processes agree on different sets. While the fault on the rank 5 process does not affect the result correctness, the one on rank 2 separates the rank 3 process from the rest. We could prevent this behaviour by re-assigning the duties of the failed process to another non-failed one.

Following the above concept, we update Algorithm~\ref{alg:tree} to consider the duties re-assignment. We can use the \texttt{MPI\_Recv} to detect the failure of a process since it will either wait for a result or show an error. We already use the receive function in the base version of our algorithm: we can control for the neighbours' liveness without modifications. Given this observation, we can impose the failed process duties to move to the next non-failed one. This assumption ensures that fallback routes exist in our algorithm, preventing partitions. The fallback selection is also unequivocal, meaning that it is unique, and all the processes detect the same without communicating with each other. Using these remarks, we can define the behaviour upon noticing a fault. In particular, a process would try to contact the successors of the failed one individually until receiving a response. If all the successors between the process and the failed one do not respond, it assumes to be the closest live successor, so it heirs the failed process duties.

\begin{figure}[!t]
    \centering
    \includegraphics[width=0.5\textwidth]{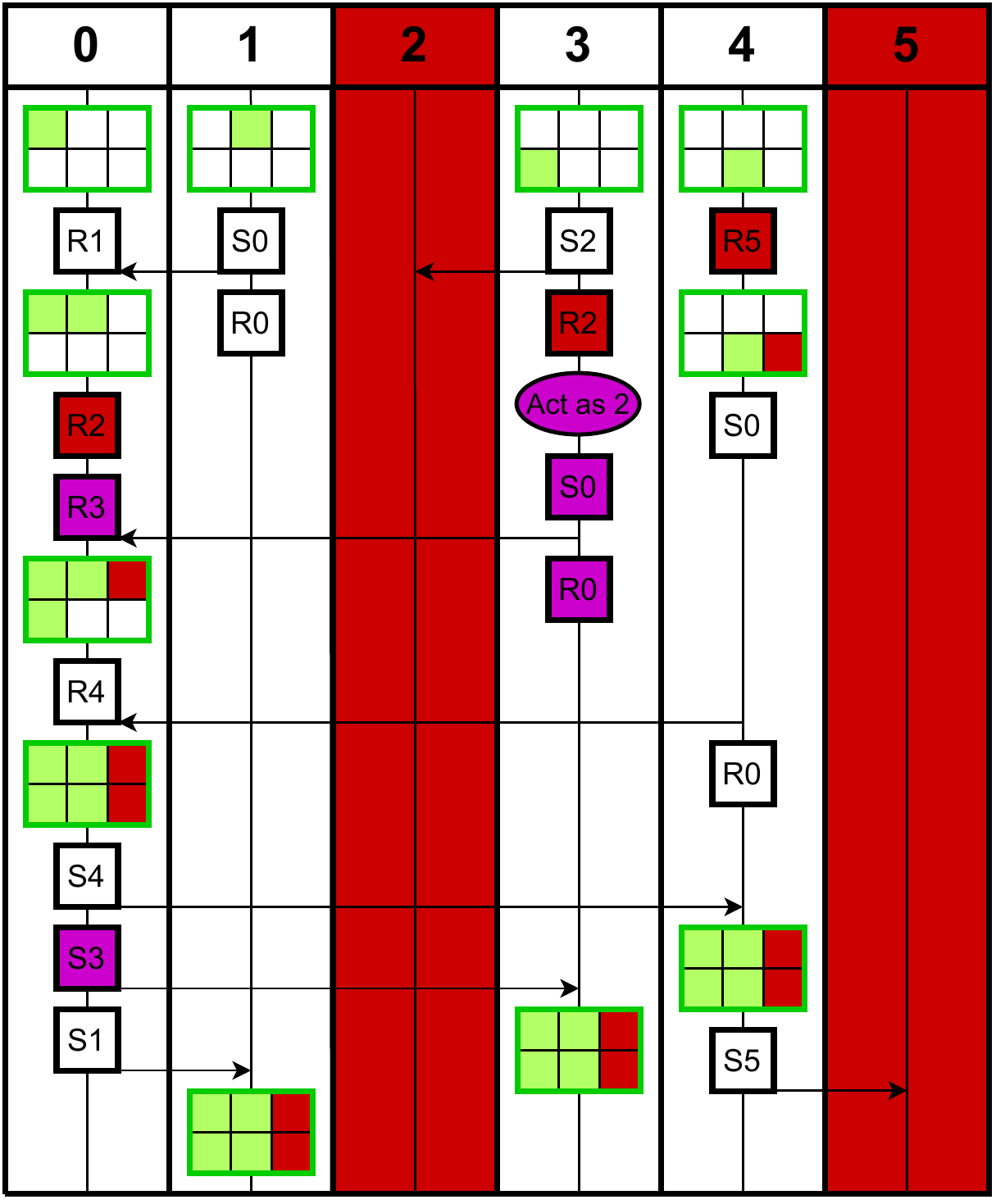}
    \caption{The figure shows the algorithm adaptation in presence of faults. Purple shapes represent additional or different calls due to the adaptation.}
    \label{fig:tree_bypass}
\end{figure}

Figure~\ref{fig:tree_bypass} shows the behaviour of the updated algorithm in the presence of faults. It is possible to see that the execution outcome is correct despite failed processes because rank 3 gets the failed rank duties. Our changes affect the algorithm complexity: the worst case goes from logarithmic to linear complexity due to the single checks of all the failed process successors.

With this algorithm, all the processes can adjust the group parameter to remove failed ones. If we use this algorithm, the two non-collective calls do not manifest the deadlock eventuality. In particular, the \texttt{MPI\_Comm\_create\_group} function exposes an error while the \texttt{MPI\_Comm\_create\_from\_group} completes correctly. This result has some remarkable implications for ULFM: the existence of a call able to create a communicator despite faults' presence and without collectiveness opens the possibility for non-collective reparation. Upon detecting a failed process, processes can substitute the communicator with a new one or even partition it asynchronously. This possibility overcomes one of the main limitations of ULFM and opens new research directions in the field.

The proposed method can also remove the collectiveness constraint of the ULFM \textit{agree} call. We can use the Liveness Discovery Algorithm to perform an all-reduce operation alongside the usual calls, achieving agreement even with faults. Moreover, this procedure can happen non-collectively, removing more constraints on ULFM possibilities.

While the Liveness Discovery Algorithm solves the problems of deadlocks in non-collective communicator creation and improves the ULFM functionalities, its complexity makes it unfeasible for user-level code. Being a distributed fault-aware algorithm, we think encapsulating its complexity in an existing framework makes it easier to leverage by the users. For this reason, we integrated the Liveness Discovery Algorithm inside the Legio framework since it already leverages the PMPI profiling interface to introduce fault management support without changes in the application code. The integration with Legio allows us to call the Liveness Discovery Algorithm before any non-collective communicator creation call, remove failed processes from the groups passed as parameters and complete the functions correctly, even in a faulty scenario. 

\section{Experimental campaign}\label{sec:experimental}

The proposed experimental campaign evaluates the scalability and overhead of the proposed solutions, both in the presence and absence of faults. In particular, we focus on the scalability of the Liveness Discovery Algorithm, the overhead of the Legio integrated solution, and the comparison between ULFM \textit{shrink} and \textit{agree} functionalities with the non-collective versions. We execute our experiments on the IT4Innovations Karolina cluster, featuring nodes with 2 x AMD Zen 2 EPYC™ 7H12, 2.6 GHz processors and 256 GB of RAM. Each node can run up to 128 processes without overloading. We use the latest version of OpenMPI featuring ULFM (v5.0.0), which implements MPI standard 4.0.
In this experimental campaign, we first measure the proposed algorithm scalability and then evaluate the cost of fault management (fault discovery and remotion) in non-collective communicator creation calls. Finally, we compare the non-collective versions of the \textit{shrink} and \textit{agree} functions with their ULFM counterparts.

\begin{figure}[!t]
    \centering
    \includegraphics[width=0.7\textwidth]{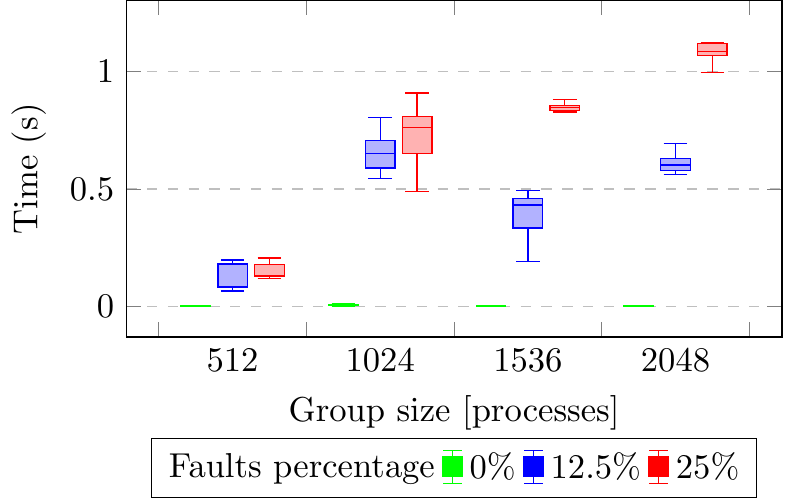}
    \caption{Execution time distribution of the Liveness Discovery Algorithm over different scenarios of group size and failure percentage.}
    \label{fig:algo_perf}
\end{figure}

With the first experiment, we evaluate the time needed to complete the Liveness Discovery Algorithm with different group sizes and different amounts of faults inside the system. We run all the experiments using 16 nodes, each with 128 processes. We choose the processes to fail randomly since their position affects the time needed to complete the algorithm. We execute the test several times for each group size and fault amount to better evaluate the results' variability. Figure~\ref{fig:algo_perf} shows the results of this evaluation. From the results, it is possible to see that the dimension of the group does not significantly affect the time needed to complete the algorithm in a fault-free scenario. Faults' presence heavily impacts the time to complete the algorithm due to the gradual shift towards linear complexity and the time to manage errors at the ULFM level. In case of no faults, the execution completes in milliseconds, showing reduced time variability. In the case of applications not using frequent communicator creation functions, we think that the observed overhead is negligible.

\begin{figure}[!t]
    \centering
    \includegraphics[width=0.9\textwidth]{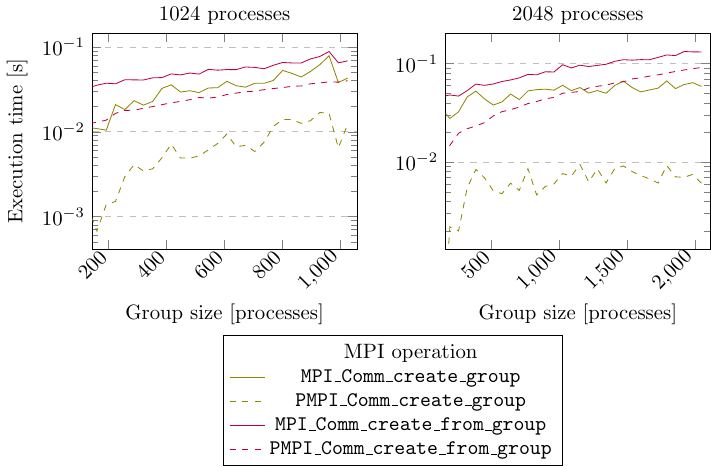}
    \caption{Mean execution time of the Legio-integrated non-collective operations (\texttt{MPI\_*}) compared to the original fault unaware versions (\texttt{PMPI\_*}).}
    \label{fig:overhead_evo}
\end{figure}

\begin{figure}[!t]
    \centering
    \includegraphics[width=0.9\textwidth]{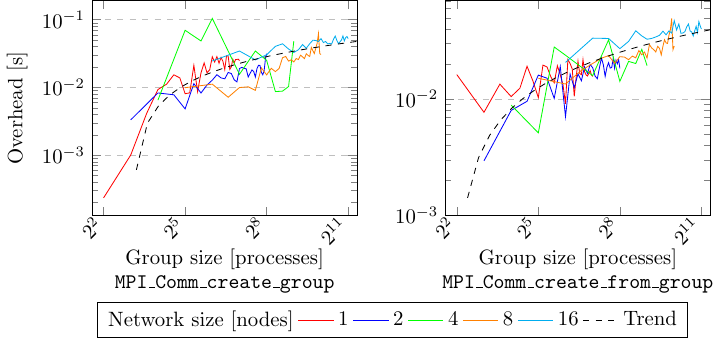}
    \caption{The figure shows the correlation of the non-collective function overhead with the group size and number of nodes. The dashed black line represents the logarithmic trend followed by the measurements.}
    \label{fig:overhead}
\end{figure}

With the second experiment, we measure the overhead introduced in the non-collective communication calls in a fault-free scenario. We execute the benchmark several times, measuring the average time needed to complete the calls. We compare the results with the time to complete the function without the additional fault discovery and removal functionalities. We repeat the experiment over networks of different sizes and with variable group dimensions to evaluate the scalability of our proposed integration. Figure~\ref{fig:overhead_evo} shows the evolution of execution times with networks of 1024 and 2048 processes (8 and 16 nodes), while Figure~\ref{fig:overhead} compares the overhead observed per function over different network sizes. The results show that the group size influences the overhead more than the network size, and the overhead follows a logarithmic trend. These considerations prove the scalability of the proposed integration in a fault-free scenario.

\begin{figure}[!t]
    \centering
    \includegraphics[width=\textwidth]{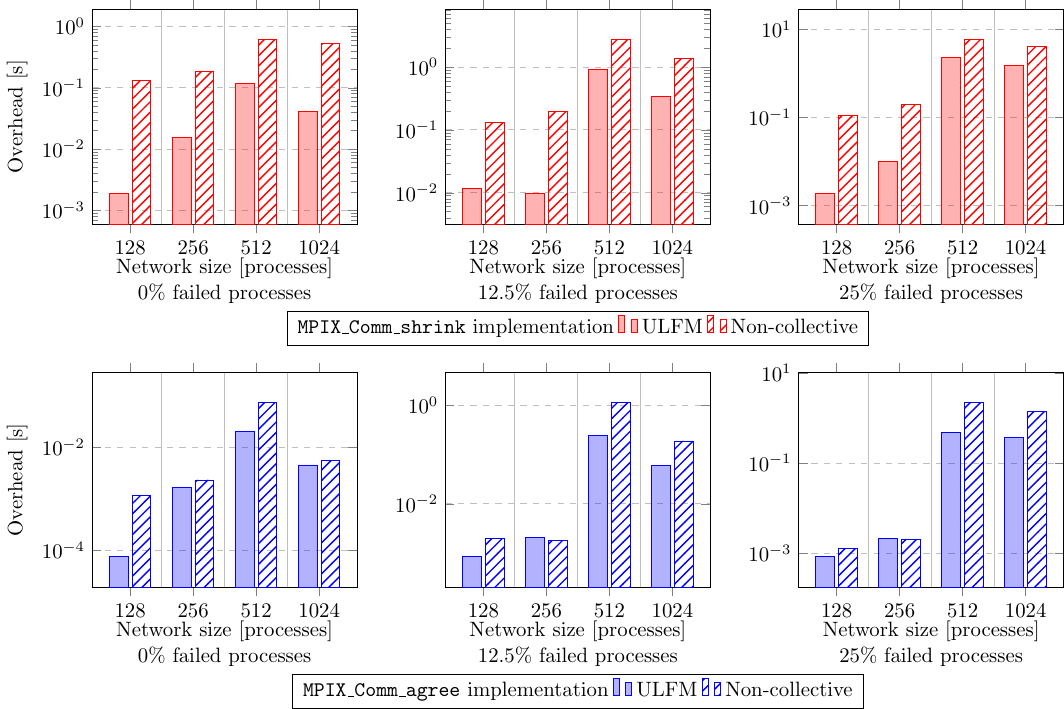}
    \caption{The figure compares the mean execution times of the non-collective reparation functions with their ULFM counterpart over different network and failure sizes.}
    \label{fig:ulfm_compare}
\end{figure}

With the last experiment, we evaluate the performance of the proposed non-collective alternatives of the ULFM functions \textit{shrink} and \textit{agree}. We compare the execution times over networks of different sizes (from 1 to 16 nodes) and with various amounts of faults. We repeat each experiment 10 times and extract the mean time to complete the functions. Figure~\ref{fig:ulfm_compare} shows the execution time comparison. The proposed non-collective alternatives require a little more execution time than their ULFM counterparts. The additional time is noticeable in the \textit{shrink} operation, while the \textit{agree} call performs similarly to its ULFM counterpart. These experiments compared the performance in a collective scenario, which should further benefit the ULFM approach. The results obtained validate the proposed non-collective ULFM alternatives, making them a promising addition to the ULFM proposal.

\section{Conclusions}\label{sec:conclusion}

In this effort, we presented a methodology to manage faults in a non-collective way. The proposed solution is mandatory for applications leveraging non-collective communicator creation functions, which would incur unnecessary synchronizations using only ULFM functionalities. The designed Liveness Discovery Algorithm proactively detects failed processes, allowing their remotion from the communicator creation and the subsequent completion of the call. 

The experimental campaign showed the effectiveness of the design, development and integration of the Liveness Discovery Algorithm with low overhead. Its usage with ULFM enables non-collective communication creation and opens for non-collective communicator reparation. We also showed that the non-collective variants of the ULFM \textit{shrink} and \textit{agree} have a manageable overhead compared to the collective ones. This result is relevant since it removes one of the weaknesses of ULFM, making it more appealing for applications.

Nonetheless, we think that our proposed Liveness Discovery Algorithm is not feasible for direct integration inside user application code. Its complexity would make the code difficult to read and harder to maintain. Moreover, the algorithm must execute before the non-collective call, unlike most ULFM functions. This fact implies that the usual packaging of ULFM functionalities inside error handlers is not feasible for this case. We integrated the algorithm inside the Legio library to leverage it transparently in our applications, but we think a ULFM extension is also possible.


\bibliography{samplepaper}

\begin{thebibliography}{10}
\providecommand{\url}[1]{\texttt{#1}}
\providecommand{\urlprefix}{URL }
\providecommand{\doi}[1]{https://doi.org/#1}

\bibitem{Top500}
Top500, the list of the most powerful hpc supercomputers,
  \url{https://www.top500.org/lists/top500/2022/11/}

\bibitem{bland2013post}
Bland, W., Bouteiller, A., Herault, T., Bosilca, G., Dongarra, J.: Post-failure
  recovery of mpi communication capability: Design and rationale. The
  International Journal of High Performance Computing Applications
  \textbf{27}(3),  244--254 (2013)

\bibitem{bouteiller2022implicit}
Bouteiller, A., Bosilca, G.: Implicit actions and non-blocking failure recovery
  with mpi. arXiv preprint arXiv:2212.08755  (2022)

\bibitem{bouteiller2006mpich}
Bouteiller, A., Herault, T., Krawezik, G., Lemarinier, P., Cappello, F.:
  Mpich-v project: A multiprotocol automatic fault-tolerant mpi. The
  International Journal of High Performance Computing Applications
  \textbf{20}(3),  319--333 (2006)

\bibitem{clarke1994mpi}
Clarke, L., Glendinning, I., Hempel, R.: The mpi message passing interface
  standard. In: Programming environments for massively parallel distributed
  systems, pp. 213--218. Springer (1994)

\bibitem{dinan2011noncollective}
Dinan, J., Krishnamoorthy, S., Balaji, P., Hammond, J.R., Krishnan, M.,
  Tipparaju, V., Vishnu, A.: Noncollective communicator creation in mpi. In:
  European MPI Users' Group Meeting. pp. 282--291. Springer (2011)

\bibitem{egwutuoha2013survey}
Egwutuoha, I.P., Levy, D., Selic, B., Chen, S.: A survey of fault tolerance
  mechanisms and checkpoint/restart implementations for high performance
  computing systems. The Journal of Supercomputing  \textbf{65}(3),  1302--1326
  (2013)

\bibitem{fagg2000ft}
Fagg, G.E., Dongarra, J.J.: Ft-mpi: Fault tolerant mpi, supporting dynamic
  applications in a dynamic world. In: European Parallel Virtual
  Machine/Message Passing Interface Users’ Group Meeting. pp. 346--353.
  Springer (2000)

\bibitem{ferreira2011rmpi}
Ferreira, K., Riesen, R., Oldfield, R., Stearley, J., Laros, J., Pedretti, K.,
  Brightwell, R.: rmpi: increasing fault resiliency in a message-passing
  environment. Sandia National Laboratories, Albuquerque, NM, Tech. Rep.
  SAND2011-2488  (2011)

\bibitem{gamell2014exploring}
Gamell, M., Katz, D.S., Kolla, H., Chen, J., Klasky, S., Parashar, M.:
  Exploring automatic, online failure recovery for scientific applications at
  extreme scales. In: SC'14: Proceedings of the International Conference for
  High Performance Computing, Networking, Storage and Analysis. pp. 895--906.
  IEEE (2014)

\bibitem{gamell2015local}
Gamell, M., Teranishi, K., Heroux, M.A., Mayo, J., Kolla, H., Chen, J.,
  Parashar, M.: Local recovery and failure masking for stencil-based
  applications at extreme scales. In: SC'15: Proceedings of the International
  Conference for High Performance Computing, Networking, Storage and Analysis.
  pp. 1--12. IEEE (2015)

\bibitem{losada2019local}
Losada, N., Bosilca, G., Bouteiller, A., Gonz{\'a}lez, P., Mart{\'\i}n, M.J.:
  Local rollback for resilient mpi applications with application-level
  checkpointing and message logging. Future Generation Computer Systems
  \textbf{91},  450--464 (2019)

\bibitem{losada2017resilient}
Losada, N., Cores, I., Mart{\'\i}n, M.J., Gonz{\'a}lez, P.: Resilient mpi
  applications using an application-level checkpointing framework and ulfm. The
  Journal of Supercomputing  \textbf{73}(1),  100--113 (2017)

\bibitem{pauli2015intrinsic}
Pauli, S., Arbenz, P., Schwab, C.: Intrinsic fault tolerance of multilevel
  monte carlo methods. Journal of Parallel and Distributed Computing
  \textbf{84},  24--36 (2015)

\bibitem{rocco2021legio}
Rocco, R., Gadioli, D., Palermo, G.: Legio: fault resiliency for embarrassingly
  parallel mpi applications. The Journal of Supercomputing pp. 1--21 (2021)

\bibitem{shahzad2018craft}
Shahzad, F., Thies, J., Kreutzer, M., Zeiser, T., Hager, G., Wellein, G.:
  Craft: A library for easier application-level checkpoint/restart and
  automatic fault tolerance. IEEE Transactions on Parallel and Distributed
  Systems  \textbf{30}(3),  501--514 (2018)

\bibitem{suo2013nr}
Suo, G., Lu, Y., Liao, X., Xie, M., Cao, H.: Nr-mpi: a non-stop and fault
  resilient mpi. In: 2013 International Conference on Parallel and Distributed
  Systems. pp. 190--199. IEEE (2013)

\bibitem{teranishi2014toward}
Teranishi, K., Heroux, M.A.: Toward local failure local recovery resilience
  model using mpi-ulfm. In: Proceedings of the 21st european mpi users' group
  meeting. pp. 51--56 (2014)

\end{thebibliography}

\end{document}